\documentclass[prb,twocolumn,superscriptaddress,longbibliography]{revtex4-1}

\usepackage{amsmath}
\usepackage{amssymb}
\usepackage{amsthm}

\usepackage{graphicx}
\usepackage{graphics}
\usepackage{bbold,bm}
\usepackage{hyperref}
\usepackage{epsfig}
\usepackage{dcolumn}

\usepackage{color}

\newcommand\vex[1]{\mathbf{#1}}
\newcommand\gvex[1]{\boldsymbol{#1}}

\def\sgn{\mathrm{sgn}}
\def\re{\mathrm{Re}\,}

\def\id{\mathbb{1}}

\def\trans{\mathsf{T}}

\def\dbar{\hbox{$d$\kern-0.6em\raise0.3em\hbox{$-$}}\hspace{-0.5mm}}

\begin{document}

\title{Surface spectra of Weyl semimetals through self-adjoint extensions}

\author{Babak Seradjeh}
\affiliation{Department of Physics, Indiana University, Bloomington, Indiana 47405, USA}
\affiliation{Max Planck Institute for the Physics of Complex Systems, N\"othnitzer Stra\ss e 38, 01187 Dresden Germany}

\author{Michael Vennettilli} 
\affiliation{Department of Physics and Astronomy, Purdue University, West Lafayette, Indiana 47907, USA}
\affiliation{Department of Physics and Astronomy, Ursinus College, Collegeville, Pennsylvania 19426, USA}


\begin{abstract}
We apply the method of self-adjoint extensions of Hermitian operators to the low-energy, continuum Hamiltonians of Weyl semimetals in bounded geometries and derive the spectrum of the surface states on the boundary. This allows for the full  characterization of boundary conditions and the surface spectra on surfaces both normal to the Weyl node separation as well as parallel to it. We show that the boundary conditions for quadratic bulk dispersions are, in general, specified by a $\mathbb{U}(2)$ matrix relating the wavefunction and its derivatives normal to the surface. We give a general procedure to obtain the surface spectra from these boundary conditions and derive them in specific cases of bulk dispersion. We consider the role of global symmetries in the boundary conditions and their effect on the surface spectrum. We point out several interesting features of the surface spectra for different choices of boundary conditions, such as a Mexican-hat shaped dispersion on the surface normal to Weyl node separation. We find that the existence of bound states, Fermi arcs, and the shape of their dispersion, depend on the choice of boundary conditions. This illustrates the importance of the physics at and near the boundaries in the general statement of bulk-boundary correspondence.
\end{abstract}

\maketitle

\section{Introduction}
Topological phases of matter have garnered an increasingly central standing in condensed matter physics and related fields. A generic feature of such phases is an intimate correspondence between nontrivial topology of bulk states, expressed in terms of topological invariants of bulk bands under periodic boundary conditions, and the appearance of robust states bound to the physical edges of the system in a bounded geometry. Examples of this \emph{bulk-boundary correspondence} abound: quantum hall phases in two dimensions support chiral edge modes;~\cite{Lau81a,ThoKohNig82a} time-reversal invariant topological phases in two and three dimensions support gapless edge and surface states;~\cite{KanMel05a,KanMel05b,WuBerZha06a,FuKanMel07a,MooBal07a,Roy06a} topological superconductors support Majorana modes at edges, surfaces, and inside vortex cores;~\cite{Kit01a,FuKan08a,SerGro11a,QiZha11a} and Weyl semimetals are known to support unconventional surface states with open ``Fermi arcs.''~\cite{BurBal11a,WanTurVis11a,HalBal12a} Intuitively, the appearance of these robust surface states regardless of local details of the surface is the result of the global, topological character of the bulk.

Despite many examples and its intuitive appeal, a precise theory of the bulk-boundary correspondence is lacking.~\cite{LiuZhoWan17a,SedKalDut17a,RhiBarSla17a}  Surface states are, of course, not special to topological phases. Tamm~\cite{Tam32a} and then Shockley~\cite{Sho39b} showed a long time ago that terminating a periodic potential generically leads to the appearance of states bound to the termination surface, whose energy and wavefunction depend on the boundary conditions at the surface. The role of boundary conditions for surface states of topological phases is, however, not systematically studied. The form of the boundary condition is often not specified or assumed ad hoc.~\cite{TeoKan10a,OkuMur14a} In many cases, the bulk Hamiltonian is stripped down to a linear bulk dispersion,~\cite{DevVol17a} which would then only support a restricted set of boundary conditions, say, characterized by a single parameter,~\cite{LiAnd15a,HasKimWu17a} which artificially limits the range of possible physical effects at the surface. A recent work~\cite{KhaMayHan17a} considers a more general dispersion; however, as our approach illustrates, the conditions imposed for the boundary conditions in this work are also too restrictive; as a result, not all physically possible boundary conditions are accounted.

In this work, we address this question by studying the effect of boundary conditions on surface spectra of Weyl semimetals in a bounded geometry in the most general terms. We do so by studying the generic low-energy, continuum Hamiltonians of a Weyl semimetal and their self-adjoint extensions in bounded geometries.~\cite{Neu29a,GitTuyVor12a,AhaOrtSer16a} This approach allows us to find all physically possible boundary conditions for the given continuum Hamiltonian. Further restrictions on the boundary conditions can be imposed by symmetries of the bulk Hamiltonian. However, it must be emphasized that respecting those symmetries is not necessary at the boundary, where bulk symmetries may be broken extraneously or spontaneously.~\cite{Cap77a} While not every boundary condition is easily realized or practical in the lab, each boundary condition represents physics that is, in principle, possible at and near the boundary.

Specifically, we consider two generic Hamiltonians supporting a pair of degenerate Weyl nodes: one with a minimal bulk dispersion, and the other with an isotropic bulk dispersion. We show that the boundary conditions are in general specified by a $\mathbb{U}(2)$ matrix relating the wavefunction and its derivatives normal to the surface.  On a surface parallel to the Weyl node separation, we find conditions for the existence of Fermi arcs. In particular, for the minimal dispersion, we fully characterize these conditions in terms of a $\mathbb{U}(1)$ phase. For the isotropic dispersion, we find the subset of $\mathbb{U}(2)$ boundary conditions that support straight-line Fermi arcs connecting the Weyl nodes on the surface. In both cases, all self-adjoint boundary conditions spontaneously break particle-hole symmetry on these surfaces. Moreover, we show that with the simplest boundary condition where the wavefunction vanishes on the surface, the surface spectrum is linear, with surface bound states whose confinement length is independent of the dispersion and a standing-wave modulation normal to the surface. Moreover, for the surface normal to the Weyl node separation, we characterize the surface bound state spectrum fully in the case of the minimal dispersion and give specific results for the isotropic dispersion. For the minimal dispersion, the energy bands are those of a gapped Dirac spectrum. Interestingly, for the minimal dispersion, the surface states only exist over a range of momenta determined by spin-orbit coupling, and their energy bands are gapped in the shape of a Mexican hat. This raises the possibility of realizing exotic electronic phases at the surface in the presence of interactions or by local potential engineering.


\section{Model}
For a Weyl semimetal with nodes separated by a wave vector $k_0$ much smaller than the inverse lattice spacing, we can take a minimal continuum model with Bloch Hamiltonian
\begin{equation}\label{eq:Hk}
H_{\vex k} = \begin{bmatrix} 
  \epsilon_{\vex{k}}     & \lambda(k_x-ik_y)\\ 
  \lambda(k_x+ik_y) & -\epsilon_{\vex{k}} 
\end{bmatrix} \equiv \vex d_{\vex k}\cdot\gvex\sigma,
\end{equation}
where $\vex k =(k_x,k_y,k_z)$ is the crystal momentum (we use natural units $\hbar=1$), $\gvex\sigma$ is the vector of Pauli matrices, $\vex d_{\vex k}=(\lambda k_x,\lambda k_y,\epsilon_{\vex k})$, $\lambda$ parameterizes the spin-orbit interaction, and $\epsilon_{\vex k}$ is the bulk dispersion relation in the absence of spin-orbit coupling. We shall consider two dispersion relations: (1) $\epsilon_{\vex k} = (k_z^2-k_0^2)/2m$ corresponds to the minimal model of Ref.~\onlinecite{OkuMur14a}; and (2) $\epsilon_{\vex k} = (k^2-k_0^2)/2m$ is isotropic in $\vex k$. Here $k_0, m>0$ are constants. 

The bulk spectrum is given by $\pm E_{\vex k}$,
\begin{equation}
E_{\vex k} = |\vex d_{\vex k}| = \sqrt{\epsilon_{\vex k}^2+\lambda^2(k_x^2+k_y^2)}.
\end{equation}
There are two Weyl nodes at $\pm \vex k_0$ with $\vex k_0 = (0,0,k_0)$. The winding of the vector $\vex d_{\vex k}$ in the Brillouin zone determines the existence of the Weyl nodes through the monopoles of the Berry's flux. The monopole charge is given by 
\begin{equation}
\sgn \det \frac{\partial \vex d_{\vex k}}{\partial \vex k}\bigg\vert_{\vex k = \pm\vex k_0} = \pm 1.
\end{equation}
In the following, we will work with dimensionless quantities, $\vex k/k_0 \to \vex k$, $2m\lambda/k_0 \to \lambda$, and $2m E/k_0^2 \to E$. Then the two dispersions take the form: (1) $\epsilon_{\vex k} = k_z^2-1$; and (2) $\epsilon_{\vex k} = k^2-1$.


We now note some of the symmetries of the bulk model:
\begin{enumerate}
\item There is an inversion symmetry represented by the unitary $\Pi = \sigma_z$:
\begin{equation}
\Pi H_{\vex k} \Pi = H_{-\vex k}.
\end{equation}
We note that this is not the same as the parity in the full system, which is broken in the Weyl semimetal. This symmetry describes the symmetric splitting of Weyl nodes in the minimal model and is therefore only approximate in a local neighborhood around the pair of Weyl nodes in the Brillouin zone.
\item There is an anti-unitary particle-hole symmetry represented by $\Gamma = \sigma_y K$, $\Gamma^2 = -\id$, where $K$ is the complex conjugation: 
\begin{equation}
\Gamma H_{\vex k} \Gamma^{-1} = \sigma_y H_{\vex k}^* \sigma_y = - H_{\vex k}.
\end{equation}
For an eignestate $\psi_E$ of energy $E$, $\Gamma\psi_E$ is an eigenstate of energy $-E$. We see that for $E=0$ the Hamiltonian commutes with $\Gamma$; then, since $\Gamma^2=-\id$, by Kramer's theorem the zero-energy subspace is doubly degenerate. We also note that the operator $\Theta=\Gamma\Pi$, $\Theta^2=\id$ is an effective time-reversal operator, $\Theta H_{\vex k}\Theta^{-1} = H_{-\vex k}$.
\item Reflection symmetries through the plane normal to the $x^\mu$-direction, $x^\mu=x,y,z$, are represented by $\Upsilon^\mu = (\sigma_x,\sigma_y,\id)$:
\begin{equation}
\Upsilon^\mu H_{\vex k} \Upsilon^\mu = (-1)^{\rho_\mu} H_{\mathcal{R}^\mu\vex k},
\end{equation}
where the sign $(-1)^{\rho_\mu} = (-1,-1,1)$ and $\mathcal{R}^\mu{\vex k}$ is the reflection of $\vex k$. Defining $\vex k = \vex k^\mu_\parallel + \vex k^\mu_\perp$ with parallel and perpendicular components to the plane, $\mathcal{R}^\mu\vex k = \vex k^\mu_\parallel - \vex k^\mu_\perp$. Thus, our model Hamiltonian is even under a reflection through the $xy$-plane, and odd under reflections through $xz$- and $yz$-planes, since the spin-orbit term breaks the latter symmetries. However, the two broken reflection symmetries result in a spectral symmetry, whereby an eigenstate $\psi_{E}$ of energy $E$ is mapped to eigenstates $\Upsilon^x\psi_E$ and $\Upsilon^y\psi_E$ of energy $-E$. Note that $\Upsilon^x\Upsilon^y=i\Pi$.
\end{enumerate}
On the surface normal to the Weyl node separation, all reflection symmetries are broken; thus, we only focus on the effective particle-hole symmetry. On surfaces parallel to the Weyl node separation, the reflection symmetry through the surface normal to the Weyl node separation is respected. In our model, this is represented in the spinor space by the identity matrix and a sign reversal of $k_z$; thus, as we shall see, our results below respect this symmetry trivially. We shall also see that the effective particle-hole and time-reversal symmetries on these surfaces are spontaneously broken for all self-adjoint extensions of the bulk Hamiltonian.


\section{Surface bound states through self-adjoint extension}
In order to study the surface states, we consider a semi-infinite geometry terminating at a single two-dimensional plane normal to the $x^\mu$-direction. The momentum $\vex k^\mu_\parallel$ parallel to the plan remains conserved, but the momentum normal to the plane must be treated as the normal component, $\hat{\vex p}^\mu_\perp$, of the momentum operator $\hat{\vex p} \to -i\gvex\nabla$ in the position basis representation. Correspondingly, the Hamiltonian is the operator $\hat H^\mu=H_{\vex k^\mu_\parallel + \hat{\vex p}^\mu_\perp}$. The presence of the surface affects the spatial symmetries of the Hamiltonian.

In a bounded or semi-bounded geometry, the specification of boundary conditions is an integral part of the definition of a self-adjoint operator: different boundary conditions result in different spectral properties. 
This is because at the physical boundary of the system, different boundary conditions specify different physics at the interface, say with a trivial or topological insulator, a regular or topological metal or semimetal, a superconductor, etc. In this work, we take an agnostic view of what lies on the other side of the interface. Instead, we would like to specify all possible boundary conditions physically admissible. 
We also pay attention to interfaces that preserve certain bulk symmetries of the system. However, we note that from a physical point of view, symmetry-breaking boundary conditions are no less interesting, since they can describe interfaces with material that break the corresponding symmetries at the surface.

In a semi-bounded geometry, in order for the Hamiltonian $\hat H^\mu$ to be self-adjoint, its domain $\mathcal{D}(\hat H^\mu)$ in the Hilbert space needs to be properly defined. 
The set of self-adjoint operators is determined fully by von Neumann's deficiency index theorem.~\cite{AhaOrtSer16a} This method, while comprehensive, is technical and not physically transparent. Instead, we use a formulation based on a corresponding conserved current derived from the unitary evolution generated by the self-adjoint operator. For two states $\psi, \phi \in \mathcal{D}(\hat H^\mu)$, the current defined as
\begin{equation}\label{eq:SAEj}
j^\mu = i \int^{x^\mu}_{x_0} \left[\psi^\dag(\hat H^\mu \phi) - (\hat H^\mu\psi)^\dag \phi\right]\hspace{-1.15mm}(\bar x^\mu)\: d\bar x^\mu + j_0^\mu,
\end{equation}
where $x_0$ is an arbitrary reference point and $j_0^\mu=j^\mu(x_0)$, is local and conserved: it satisfies the continuity equation
\begin{equation}
\frac{\partial \rho}{\partial t}+\frac{\partial j^\mu}{\partial x^\mu}=0,
\end{equation}
with $\rho = \psi^*\phi$. By unitarity of time evolution, $\frac{d}{d t}\int \rho\: dx = 0$, where the integral is over the entire system. Since in a semi-infinite geometry, the wave functions vanish at infinity, we must have $j^\mu(\infty)=0$. Thus,
\begin{equation}\label{eq:SAEj}
j^\mu(0)=0.
\end{equation}
This conserved current is a sesquilinear, Hermitian form of the two arbitrary states $\psi$ and $\phi$. In general, Eq.~(\ref{eq:SAEj}) can be expressed as $\nu^\mu_+[\psi]^\dag\nu^\mu_+[\phi] = \nu^\mu_-[\psi]^\dag\nu^\mu_-[\phi]$, where $\nu^\mu_\pm[\psi]$ is, for our quadratic two-band Hamiltonians, a 2-component vector of the wave function and its derivative, normal to the surface, at $x^\mu=0$. In turn, this yields the boundary condition
\begin{equation}
\nu^\mu_+[\psi] = U \nu^\mu_-[\psi]
\end{equation}
where the unitary matrix $U\in\mathbb{U}(2)$ parametrizes the boundary conditions. We summarize the relevant results in the Appendix.

In the following, we parametrize an element of $\mathbb{U}(2)$ as
\begin{equation}\label{eq:U2}
U = e^{im_0}\begin{bmatrix}
e^{i\alpha}\cos\delta  & -e^{-i\beta}\sin\delta \\
e^{i\beta}\sin\delta  & e^{-i\alpha}\cos\delta 
\end{bmatrix},
\end{equation}
As a special case, we note the subset with $\sin\delta=0$ corresponds to a diagonal $\mathbb{U}(1)\times\mathbb{U}(1)$ subgroup. 

The condition to find the spectrum of surface bound states can be obtained by writing the bound state wavefunction $e^{i \vex k^\mu_\parallel\cdot \vex x ^\mu_\parallel} \psi_b^\mu(\vex k^\mu_\parallel, x^\mu)$, with $\vex x^\mu_\parallel$ the coordinates on the surface, and the bound state
\begin{equation}\label{eq:psibu}
\psi_b^\mu(\vex k^\mu_\parallel, x^\mu) = \frac1{\sqrt{\kappa^\mu}}e^{-\kappa^\mu x^\mu}\psi^\mu_0,
\end{equation}
where $\kappa^\mu$ with $\re\kappa^\mu>0$ is the dimensionless inverse confinement length, which may also depend on $\vex k^\mu_\parallel$. The energy of the bound state is $\pm E_b^\mu$,
\begin{equation}\label{eq:Ebu}
E_b^\mu (\vex k^\mu_\parallel, \kappa^\mu) = E_{\gvex\kappa^\mu},
\end{equation}
where $\gvex\kappa^\mu={\vex k^\mu_\parallel + i\kappa^\mu {\check{\vex x}}^\mu}$ and $\check{\vex x}^\mu$ is the unit vector in the $x^\mu$-direction. The spinor $\psi_0^\mu$ has the form
\begin{equation}\label{eq:psi0u}
\psi_0^\mu \propto \begin{bmatrix} d_{x\gvex\kappa^\mu} -i d_{y\gvex\kappa^\mu} \\ E_b^\mu - d_{z\gvex\kappa^\mu}\end{bmatrix}
\end{equation}
up to a normalization factor.

In order to determine $\kappa^\mu$ and $\psi_0^\mu$ for a given value of energy $E_b^\mu = E$, we invert Eq.~(\ref{eq:Ebu}) and write $\kappa^\mu$ as a function of $E$. There are two cases that we treat separately:
\begin{enumerate}
\item If there is only one solution with $\re\kappa^\mu>0$, the corresponding bound state in Eq.~(\ref{eq:psibu}) must satisfy the boundary conditions, which we can write as
\begin{equation}
N^\mu_+ \psi_0^\mu = UN^\mu_- \psi_0^\mu,
\end{equation}
where
\begin{equation}
N^\mu_\pm \psi_0^\mu = \nu^\mu_\pm[\psi_b]
\end{equation}
depend on $\kappa^\mu$. The condition to find a nontrivial solution to $\psi_0^\mu$ is then
\begin{equation}\label{eq:NUN}
\det(N^\mu_+-UN^\mu_-)=0.
\end{equation}
\item If there are more than one possible solutions $\psi^\mu_{bn}$ with $\re\kappa^\mu_n>0$, indexed by $n$, we write a general solution as a superposition,
\begin{equation}\label{eq:superpsibu}
\psi_b^\mu = \sum_{n} c_n \psi^\mu_{bn},
\end{equation}
which must satisfy the boundary conditions, which we can write as
\begin{equation}
M^\mu_+ c = UM^\mu_- c,
\end{equation}
where $c=(c_1, c_2, \cdots)^\trans$, and
\begin{equation}
M^\mu_\pm c = \nu^\mu_\pm[\psi_b^\mu].
\end{equation}
This is then solved for $E$ and $c$ with the condition for solutions to exists given by
\begin{equation}\label{eq:MUM}
\det(M^\mu_+-UM^\mu_-)=0.
\end{equation}
\end{enumerate}


\section{Surface normal to Weyl node separation}
In this case, regardless of our choice of dispersion, the conserved current reads
\begin{equation}
j^z = i\left( \psi'^\dag \sigma_z \phi- \psi^\dag \sigma_z \phi' \right)
\end{equation}
where the prime stands for $d/d z$. Then, for $\psi=(\psi_1,\psi_2)^\trans$
\begin{equation}
\nu^z_\pm [\psi] = \begin{bmatrix}
\psi_1 \pm i\psi'_1 \\
\psi_2 \mp i\psi'_2
\end{bmatrix}.
\end{equation}
and,
denoting the inverse confinement length $\kappa^z=\kappa$, $N^z_\pm = \text{diag}(1\mp i\kappa, 1\pm i\kappa)$. 

For a diagonal unitary $U\in\mathbb{U}(1)\times\mathbb{U}(1)$, the boundary conditions reduce to two separate sets of conditions on each component of $\psi$ for the self-adjoint extensions of the operator $-d^2/dz^2$ over a semi-infinite geometry. The spectral properties including the surface bound states are previously studied.~\cite{BonFarVal01a} The full space of $\mathbb{U}(2)$ unitaries is much larger and, therefore, we expect a richer set of bound state spectra.

Before analyzing these spectra, let us characterize the subset of boundary conditions that preserve the relevant symmetries of the bulk. First, under particle-hole operation
\begin{equation}
\Gamma: \nu^z_\pm \mapsto \sigma_y\nu^{z*}_\pm,
\end{equation}
after which the boundary condition reads, $\nu^z_+ = \sigma_y U^* \sigma_y \nu^z_-$. Thus, the boundary condition is symmetric iff $U = \Gamma U \Gamma^{-1}$, which is satisfied for $U$ given in Eq.~(\ref{eq:U2}) with $m_0=0$ or $\pi$.

In this section we shall denote the momentum along the surface, $\vex k_\parallel^z=(k_x,k_y,0)\equiv\vex q$. We note that upon replacing $\vex k\to \gvex\kappa = \vex q+ i\kappa \check{\vex z}$ in the bulk dispersion, the resulting dispersion $E_{\gvex\kappa}$ does not necessarily represent the surface bound state spectrum. Further conditions and restrictions are imposed by the boundary conditions, from which the existence and the possible values of $\kappa$ must be extracted. However, assuming such solutions do exist, the functional form of the dispersion is given by $E_{\gvex\kappa}$.

\subsection{Minimal dispersion}
For a given energy of the bound state $E_b^z=E$  with the minimal dispersion, we find $\kappa^{2}=\pm\sqrt{E^2+\lambda^2 q^2}-1\in\mathbb{R}$, so there is at most a single solution with $\re\kappa>0$, for which $\kappa\in\mathbb{R}$ and, thus, $\kappa^{2}=\sqrt{E^2+\lambda^2 q^2}-1>0$. The bound state energy is then $\pm E_b^z$,
\begin{equation}\label{eq:minEz}
E^z_b = \sqrt{(\kappa^2+1)^2+\lambda^2q^2}.
\end{equation}
This is a gapped Dirac-like dispersion describing an insulating surface spectrum with minimum gap $\kappa^2+1$.

Now, Eq.~(\ref{eq:NUN}) yields the $\vex q$-independent solutions 
\begin{equation}\label{eq:minkz}
\kappa=\kappa^{(\pm)} = \frac{-\cos\delta\sin\alpha \pm \xi_1 \sqrt{\cos^2\delta-\cos^2m_0}}{\cos m_0+ \cos\delta\cos\alpha},
\end{equation}
where we have defined a sign $\xi_1=\sgn(\cos m_0+ \cos\delta\cos\alpha)$. Since only $\kappa\in\mathbb{R}$ is admissible, a necessary condition for the existence of surface states is
\begin{equation}\label{eq:minzness}
|\cos\delta|\geq|\cos m_0|. 
\end{equation}
Three separate cases arise, depending on the signs $\xi_1$ and $\xi_2=\sgn(\cos\delta\sin\alpha)$. First, if $\xi_1\xi_2>0$, bound states exist only when $|\cos\delta\cos\alpha|>|\cos m_0|$ and $\kappa=\kappa^{(+)}$. Second, if $\xi_1\xi_2<0$ and $|\cos\delta\cos\alpha|>|\cos m_0|$ then bound states exist and $\kappa = \kappa^{(+)}$. However, third, if $\xi_1\xi_2<0$ and $|\cos\delta\cos\alpha|<|\cos m_0|$, then bound states exist for both values of $\kappa=\kappa^{(\pm)}$. In the latter case, bound states form \emph{two} bands of surface energy bands given by Eq.~(\ref{eq:minEz})

For particle-hole-symmetric boundary conditions, $\sin m_0=0$. Then, the necessary condition~(\ref{eq:minzness}) gives also $\sin\delta=0$. These boundary conditions form the set of diagonal unitaries $U=\text{diag}(e^{i\alpha_U},e^{-i\alpha_U})$, where $e^{i\alpha_U}=\sgn(\cos\delta\cos m_0)e^{i\alpha}$. Then, bound states exist only when $\sin\alpha_U<0$, for which \emph{two degenerate} surface bands exist with $\kappa 
=-\tan(\alpha_U/2)$. 

\begin{figure}[tb]
\includegraphics[width=3.1in]{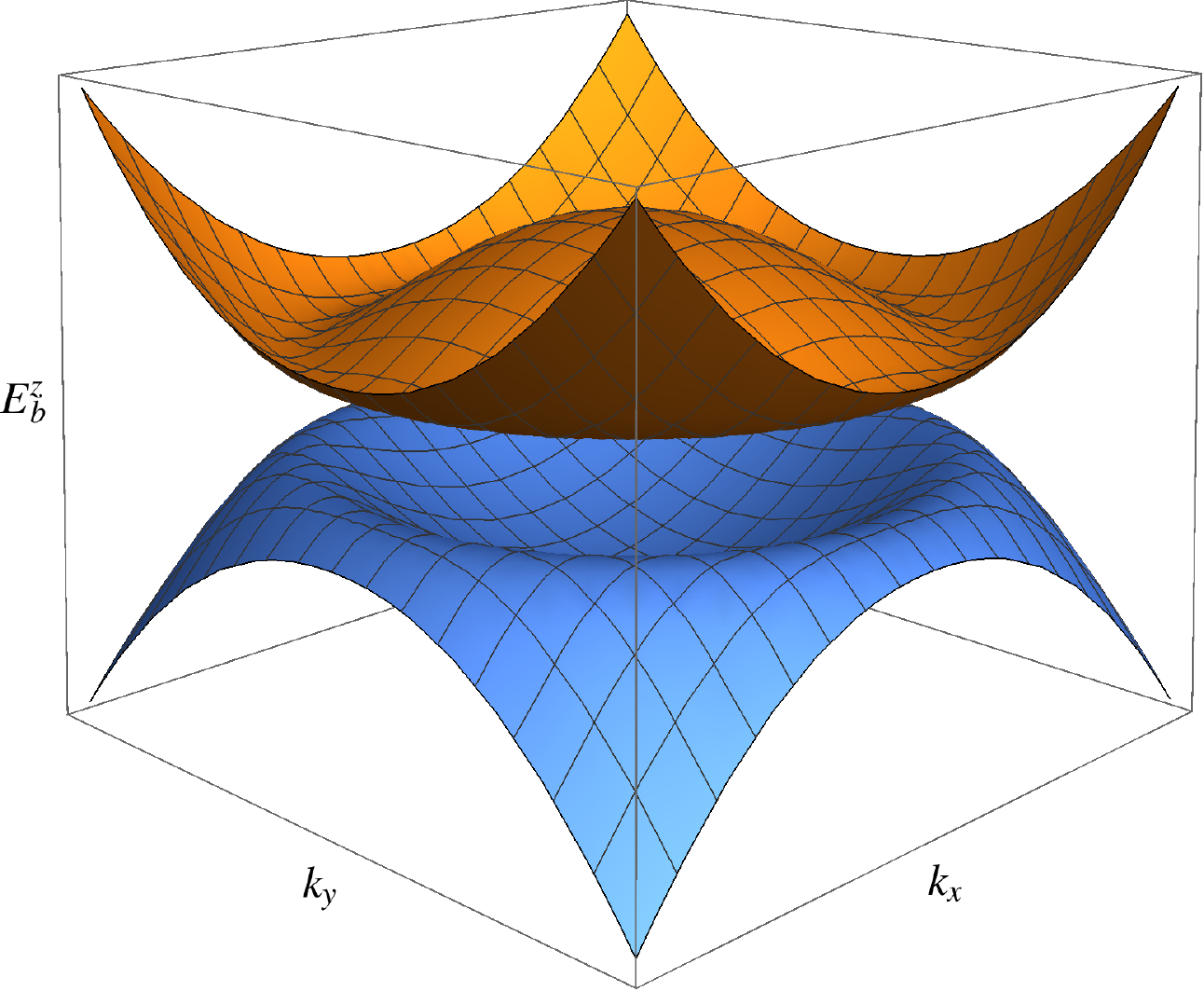}
\caption{The bound state dispersion, $E_b^z$, on the surface normal to the Weyl node separation with momentum $(k_x,k_y)$ along the surface, in units of half the separation of the Weyl nodes, and dimensionless spin-orbit coupling $\lambda^2<2(\kappa^2+1)$, where $\kappa$ is the inverse confinement length of surface bound states. 
}\label{fig:Ezb}
\end{figure}

\subsection{Isotropic dispersion}
For the isotropic dispersion, denoting $E_b^z=E$ and $\kappa^z=\kappa$, we find two possible solutions $\kappa^2_\pm = q^2-1 \pm \sqrt{E^2+\lambda^2q^2}$. When $\kappa_-^2$ becomes negative and therefore unphysical, there is only one possible value of $\kappa$ with $\re\kappa>0$. This value is found from the boundary conditions leading to Eq.~(\ref{eq:minkz}) and the subsequent analysis. A sufficient condition for $\kappa_-^2<0$ is
\begin{equation}
q < q_*=\lambda/2+\sqrt{(\lambda/2)^2+ 1}.
\end{equation}
Then, the bound state energy is $\pm E_b^z$,
\begin{equation}
E^z_b = \sqrt{(q^2-\kappa^2-1)^2+\lambda^2q^2}.
\end{equation}
This spectrum has a direct insulating gap, obtained at $q =0$ when $\lambda^2>2(\kappa^2+1)$ and  at $q_m=\sqrt{\kappa^2+1-\lambda^2/2}$ when $\lambda^2<2(\kappa^2+1)$. As shown in Fig.~\ref{fig:Ezb}, in the latter case the surface spectrum has a Mexican-hat shape 
with a minimum energy 
$
E_b^z 
	= \lambda\sqrt{q_m^2+(\lambda/2)^2}
$
at a circle with radius $q_m$ and a local maximum at $q=0$, $E_b^z = \kappa^2+1$. Thus for chemical potentials between these local minimum and maximum, the Fermi surface consists of an electron-like and a hole-like pockets. 

When both $\kappa_\pm^2>0$, we find two possible solutions with $\re\kappa_\pm>0$. In this case, taking $q_x+iq_y\equiv qe^{i\phi}$, we first find the spinor $\psi_0^z$ from Eq.~(\ref{eq:psi0u}) to be
\begin{equation}
\psi_{0\pm}^z \propto \begin{bmatrix} \lambda q e^{-i\phi} \\ E+\kappa_\pm^2+1  \end{bmatrix}.
\end{equation}
Then, taking a general superposition in Eq.~(\ref{eq:superpsibu}), we find
\begin{equation}
M_\pm =\begin{bmatrix}
\lambda q e^{-i\phi} (1\mp i\kappa_+) & \lambda q e^{-i\phi} (1\mp i\kappa_-) \\
(E+\kappa_+^2-1) (1\pm i\kappa_+) & (E+\kappa_-^2-1) (1\pm i\kappa_-)
\end{bmatrix}.
\end{equation}
Finally, $E$ is found as a solution to Eq.~(\ref{eq:MUM}).

For general $U$, the above procedure is tedious and not very instructional. Indeed, there may not even be any solutions depending on the choice of $U$, i.e. the form of the boundary conditions. To illustrate this point, let us take a simple example $U=-\id$ corresponding to the boundary condition $\psi(z=0)=0$. Then, 
\begin{align}
M_+ - U M_- &= M_++M_- \\
	&= -2i \begin{bmatrix}
	\lambda q e^{-i\phi} & \lambda q e^{-i\phi} \\
	E+\kappa_+^2-1 & E+\kappa_-^2-1
	\end{bmatrix}.
\end{align}
Thus, $\det(M_+-UM_-)=0$ only when $\kappa_+=\kappa_-$, i.e. $q=0$, which is outside the acceptable range of $q$. A similar result follows for $U=\id$, which corresponds to the boundary condition $\psi'(z=0)=0$.

This is a remarkable and unconventional result: for boundary conditions as commonly used as the vanishing of the wavefunction or its derivative on the plane, we see that not only the spectrum has an uncommon Mexican-hat shape for a range of spin-orbit couplings, but that no surface bound states exist in a ``forbidden'' range of momenta in the surface Brillouin zone.


\section{Surface parallel to the Weyl node separation: Fermi arcs}
We take the surface normal to the y-direction below and denote the momentum along the surface $\vex k_\parallel^y = (k_x,0,k_z) \equiv \vex p$.
\subsection{Minimal dispersion}
With the minimal dispersion, the conserved current takes the form
\begin{equation}
j^y = \lambda \psi^\dagger\sigma_y \phi.
\end{equation}
This current yields, in fact, $\nu_\pm[\psi] = \psi_1\pm i\psi_2$, which are complex numbers instead of spinors. Thus, the boundary conditions in this case are characterized by a single $\mathbb{U}(1)$ phase,~\cite{AhaOrtSer16a,Note1}
\begin{equation}\label{eq:minBC}
\psi_1+i\psi_2 = e^{i\gamma}(\psi_1-i\psi_2).
\end{equation}
Under particle-hole operation, 
$\Gamma: \nu_\pm \mapsto \mp\nu_\mp^*$,
and
$e^{i\gamma}\mapsto-e^{i\gamma}$.
Thus, all permissible boundary conditions supporting a self-adjoint extension of the Hamiltonian with the minimal dispersion break particle-hole symmetry on the surface parallel to Weyl node separation.~\cite{Note2}


Denoting the inverse confinement length $\kappa^y=\kappa$ and the bound state energy $E_b^y=E$, we have $E^2 = (k_z^2-1)^2 + \lambda^2(k_x^2 - \kappa^2)$. So again, for a given energy, there is at most one solution with $\re\kappa>0$. 
Solving for $\psi_0^y\propto \begin{bmatrix} 1 \\ -\rho \end{bmatrix}$ in the boundary condition equation, we have
\begin{equation}\label{eq:gamma}
\frac{1-i\rho}{1+i\rho} = e^{i\gamma} \Rightarrow \rho = \tan\frac\gamma2.
\end{equation}
Note that $\rho$  is independent of $\vex p$ and depends only on the choice of boundary condition. 
We can then find $\kappa$ from Eq.~(\ref{eq:psi0u}), which yields
\begin{equation}\label{eq:miny}
\rho = \frac{\lambda\zeta-E}{\lambda(k_x+\kappa)},
\end{equation}
with $\lambda\zeta \equiv k_z^2-1$.

\begin{figure}[tb]
\includegraphics[width=2.1in]{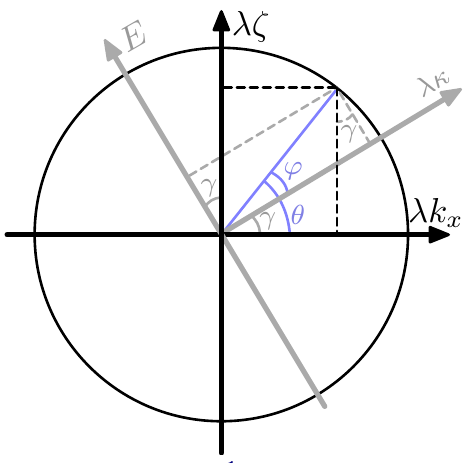}
\caption{The geometric structure of the surface state spectra normal to the $y$-direction, parallel to the direction of Weyl node separation in the $z$-direction. A pair of vectors $(\lambda\kappa, E)$ and $(\lambda k_x, \lambda\zeta)$ are related by having the same length. For the minimal dispersion, $\lambda\zeta = k_z^2-1$ and the angle $\gamma$ between these two vectors is fixed by the boundary condition. For the isotropic dispersion, there is in general two possible solutions for the inverse confinement length $\kappa$; in this case, the algebraic relationship between the boundary condition and the solutions is discussed in the text.}\label{fig:minSS}
\end{figure}

\begin{figure}[tb]
\includegraphics[width=3.5in]{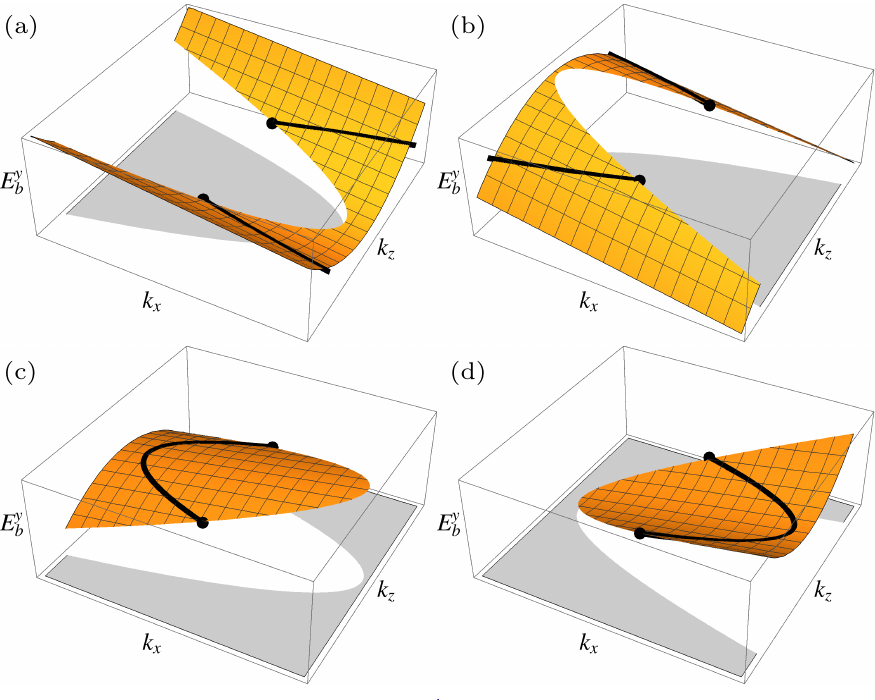}
\caption{The spectrum on the surface parallel to the Weyl node separation. The shaded region shows the ``forbidden'' region where the confinement length diverges and surface bound states are absent. Four different types of boundary conditions are characterized by the angle $\gamma$, see Eq.~(\ref{eq:gamma}), (a) $0<\gamma<\pi/2$, (b) $\pi/2<\gamma<\pi$, (c) $\pi<\gamma<3\pi/2$, and (d) $3\pi/2<\gamma<2\pi$. The black circles show the projection of the Weyl nodes on the surface and the solid black curve connecting them is the Fermi arc at $E=0$.}\label{fig:minSpec}
\end{figure}

The surface state spectrum is then found by solving Eq.~(\ref{eq:miny}). The spectrum can be rewritten as $(\lambda\kappa)^2 + E^2 = (\lambda k_x)^2 + (\lambda\zeta)^2 \equiv s^2$; defining two angular variables $-\pi<\theta<\pi$ and $-\pi/2<\varphi<
\pi/2$ as $\lambda k_x = s\cos\theta$, $\lambda\zeta = s\sin\theta$ and $\lambda\kappa = s\cos\varphi$, $E=s\sin\varphi$, we have $\tan(\gamma/2) = \rho = \tan[(\theta-\varphi)/2]$. Thus, $\theta-\varphi=\gamma$ is fixed by the boundary condition. We then find for $E=E_b^y$,
\begin{equation}
\lambda k_x = \cot\gamma\;(k_z^2-1) - \csc\gamma\; E.
\end{equation}
with the condition,
\begin{align}
\lambda \kappa  
	&= \tan\gamma\; E + \sec\gamma\;\lambda k_x \\
	&= \sin\gamma\; (k_z^2-1)+ \cos\gamma\; \lambda k_x>0.
\end{align}
The geometry of angular variables is shown in Fig.~\ref{fig:minSS}. 
This is the Fermi arc on the surface parallel to the bulk Weyl nodes. At $E=0$, it connects the Weyl nodes projected on this surface at $k_x=0,k_z=\pm 1$. At $E=0$, $\lambda \kappa = (k_z^2-1) / \sin\gamma>0$, so the Fermi arc is restricted to $(|k_z|-1)\xi_\gamma>0$, where $\xi_\gamma=\sgn\sin\gamma$: for $\xi_\gamma<0$, it stays within $|k_z|<1$; for $\xi_\gamma>0$, it extends to the Brillouin zone edges. At finite $E$, these ranges vary, dividing the Brillouin zone into allowed and ``forbidden'' regions where surface bound states are absent. These spectra are depicted in Fig.~\ref{fig:minSpec}.

Note that, in contrast to the surface normal to the Weyl node separation, the boundary conditions on the parallel surface involve both $E$ and $\kappa$. Together with the eigenvalue equation, the solutions then only allow for a single band. Thus, the bound state dispersion on this surface has a single slope, i.e. it is ``chiral.'' This is consistent with the fact that particle-hole symmetry is broken for all boundary conditions of the self-adjoint extensions of bulk Hamiltonian. (We note that, as anticipated, reflection symmetry in the $z$-direction is trivially satisfied since the dispersion is even in $k_z$.) 
This chiral dispersion reflects the topological nature of these surface states.

\subsection{Isotropic dispersion}
Now, the conserved current reads
\begin{equation}
j^y = i(\psi'^{\dagger}\sigma_z\phi - \psi^{\dagger} \sigma_z \phi') + \lambda \psi^{\dagger} \sigma_y \phi,
\end{equation}
which yields
\begin{equation}\label{eq:isonu}
\nu^y_{\pm}[\psi] = \begin{bmatrix}
\psi_1 \mp i (\psi_2 + 2\psi'_1 /\lambda ) \\
\psi_2 \pm i (\psi_1 + 2\psi'_2 /\lambda)
\end{bmatrix}.
\end{equation}
Writing the bound states $\psi_b^y \propto \begin{bmatrix} 1 \\ -\rho \end{bmatrix} e^{-\kappa y}$ with energy $E_b^y=E$, we find two possible solutions
\begin{equation}\label{eq:isokpm}
\kappa_\pm^2 = p^2  -1 +\frac12\lambda^2 \pm \sqrt{E^2+\lambda^2 [k_z^2 - 1 + (\lambda/2)^2]},
\end{equation}
and $\rho_\pm = \frac{p^2-1-\kappa_\pm^2-E}{\lambda(k_x+\kappa_\pm)}$, where $\vex p = (k_x,0,k_z)$. Note that if we define
$
\zeta_\pm = -\lambda/2 \mp \sqrt{(E/\lambda)^2+ k_z^2 - 1 + (\lambda/2)^2},
$
we have $(\lambda \kappa_\pm)^2 + E^2 = (\lambda k_x)^2 + (\lambda \zeta_\pm)^2 \equiv s_\pm^2$ and $\rho_\pm = \frac{\lambda\zeta_\pm - E}{\lambda k_x + \lambda \kappa_\pm}$. Then, for each solution the pair of vectors $(\lambda k_x, \lambda\zeta_\pm)$ and $(\lambda\kappa_\pm, E)$, with (complex) lengths $s_\pm$ and (complex) angles $\theta_\pm$ and $\varphi_\pm$ constrained to $s_+\sin\varphi_+=s_-\sin\varphi_-=E$ and $s_+\cos\theta_+=s_-\cos\theta_-=\lambda k_x$, admit a similar geometric structure as shown in Fig.~\ref{fig:minSS}.

Now, since $E^2+\lambda^2 [k_z^2 - 1 + (\lambda/2)^2]=E^2+\lambda^2 k_z^2 - \frac14\lambda^2(4-\lambda^2)\equiv W$ is not always positive, even for small $p$, we may have two solutions with $\re\kappa_\pm>0$. Indeed, for small $p$ and $E$ there is a finite range of $\lambda$ where this is true: for example, taking $E=0$ for a given $k_z$, $W<0$ for $(\lambda/2)^2<1-k_z^2$ and we find two permissible solutions for the bound states.
In this case,
\begin{equation}
M_\pm = \begin{bmatrix}
\vspace{1mm}
1 \pm i \left(\rho_+ + \frac2{\lambda}{\kappa_+}\right) & 1 \pm i \left( \rho_- + \frac2{\lambda} {\kappa_-}\right) \\
-\rho_+ \pm i \left(1 + \frac2{\lambda}{\kappa_+\rho_+}\right) & -\rho_- \pm i \left(1 + \frac2{\lambda}{\kappa_-\rho_-} \right) \\
\end{bmatrix}.
\end{equation}
In general, the spectrum can now be found from Eq.~(\ref{eq:MUM}). However, this is tedious and not very instructional. In the following we address some special cases.

First, we show that for certain boundary conditions, the Fermi arcs are straight lines connecting the Weyl nodes. In particular, assuming $|k_z|\leq 1$ and taking $E=k_x=0$, we have 
\begin{equation}
\kappa_\pm = \lambda/2 \pm \sqrt{k_z^2-1+(\lambda/2)^2} = - \zeta_\pm,
\end{equation}
and $\rho_\pm = -1$. Then,
$
M_\pm = \begin{bmatrix} 1 \pm i\omega & 1 \mp i\omega \\ 1 \mp i\omega & 1 \pm i\omega \end{bmatrix},
$
where $\omega = \frac{2}{\lambda}{\sqrt{k_z^2-1+(\lambda/2)^2}}$.
Now, taking $U$ as in Eq.~(\ref{eq:U2}) and solving Eq.~(\ref{eq:MUM}), we find
\begin{equation}\label{eq:isolinFermiArc}
\sin\delta\sin\beta = \sin m_0.
\end{equation}
For example, as opposed to the minimal dispersion, $U=-\id$, corresponding to $\psi(0)=0$, and $U=\id$, corresponding to $\psi'(0)+(\lambda/2)\sigma_x\psi(0)=0$, support straight Fermi arcs connecting the two Weyl nodes.

Next, we shall determine the spectrum for a specific choice of boundary conditions given by $U=-\id$, i.e. $\psi(0)=0$; then, $M_+-UM_- = M_++M_- = \begin{bmatrix} 2 & 2 \\ -2\rho_+ & -2\rho_- \end{bmatrix}$ and Eq.~(\ref{eq:MUM}) yields $\rho_+=\rho_-\equiv\rho$ and $c_+=-c_-$. Using the algebraic relationship defining $\rho_\pm$ and $\kappa_\pm$ we find
\begin{align}
\lambda \rho 
	&= -(\kappa_++\kappa_-), \\
E 
	&= \frac{\lambda^2}2(\rho^2-1) - \rho \lambda k_x. \label{eq:isoU1E}
\end{align}
Together with Eq.~(\ref{eq:isokpm}), these equations determine the spectrum. As must be the case, we see that for $E=k_x=0$ $\rho=-1$ and the equations are satisfied for any $|k_z|<1$ (this is necessary for having two solutions as discussed above). For small $E$ and $k_x$, we may expand $\kappa_\pm$ to find $\rho = -1 + [E^2-(\lambda k_x)^2]/[2\lambda^2(1-k_z^2)] + O(k_x^2E^2)$. Thus, up to this order, the surface state spectrum $E=E_b^y$ is given by
\begin{equation}
E_b^y = \lambda k_x.
\end{equation}
Indeed, we can show that this spectrum is correct to all orders. Taking $E=\lambda k_x$ and working backwards the two solutions $\kappa_\pm = \lambda/2\pm i\sqrt{1-(\lambda/2)^2-p^2}$, where we have assumed $p^2<1-(\lambda/2)^2$ in order to have $\re\kappa_\pm>0$. Thus, $\rho=-1$, and we see that Eq.~(\ref{eq:isoU1E}) is satisfied automatically. The wavefunction is found as
\begin{equation}\label{eq:isoz0}
\psi_b^y(\vex p,y) =
\sqrt{\lambda[1+(\lambda/2\chi)^2]}
e^{-\frac\lambda2 y}\sin\left(\chi y\right)
 \begin{bmatrix} 1 \\ 1 \end{bmatrix},
\end{equation}
with $\chi={\sqrt{1-(\lambda/2)^2-p^2}}$. Note that the surface states all have the same confinement length $2/\lambda$ and a standing-wave modulating factor with the wavenumber $\chi$  that depends on the momentum $\vex p$ on the surface. We note that particle-hole symmetry is broken for these surface states since the dispersion has a preferred chirality. 

The mapping of the general boundary conditions under particle-hole operation for the isotropic dispersion is more complicated than before: we have
\begin{equation}
\Gamma: \nu_\pm \mapsto S^{+}_{\pm}\nu_+^* + S^{-}_{\pm}\nu_-^*,
\end{equation}
with $S^\xi_\pm = \frac12(\mp\id-\xi\sigma_x+\sigma_y+i\sigma_z)$. Thus, a particle-hole symmetric boundary condition must satisfy,
\begin{align}
US^+_-U^* + US^-_- - S^+_+ U^* = S^-_+.
\end{align}
As expected from the chiral surface states we found above, it's easy to see that $U=-\id$ does not satisfy this condition: $S^+_--(S^-_--S^+_+) = \frac12(\id-\sigma_x+\sigma_y+i\sigma_z)-(\id+\sigma_x) = \frac12(-\id-3\sigma_x+\sigma_y+i\sigma_y)\neq S^-_+$. Similarly, for $U=\id$: $S^+_-+(S^-_--S^+_+) = \frac12(\id-\sigma_x+\sigma_y+i\sigma_z)+(\id+\sigma_x) = \frac12(3\id+\sigma_x+\sigma_y+i\sigma_y)\neq S^-_+$. Our numerical search for solutions to this set of unitaries returned a
null result; thus, we conclude that, as for the minimal dispersion, all self-adjoint boundary conditions of the isotropic bulk Hamiltonian also break particle-hole symmetry on the surface parallel to the Weyl node separation.

We close this section by considering the relationship between the surface bound states of the minimal and the isotropic bulk dispersions. One may expect the two cases to be related at large values of $\lambda$ where the quadratic terms in $k_x$ and $k_y$ become negligible compared to the linear spin-orbit interaction. However, this connection should be obtained at the level of boundary conditions, not the final solutions. 

For example, note that the solutions in Eq. (\ref{eq:gamma}) with the minimal dispersion and, say, Eq. (\ref{eq:isoz0}) with the isotropic dispersion have different spinor structures independent of $\lambda$. This can be seen also by the fact that the Fermi arcs of the minimal dispersion are generally curved while the Fermi arcs of the solution Eq. (\ref{eq:isoz0}) are straight lines. This is because the boundary conditions leading to Eq. (\ref{eq:gamma}) and Eq. (\ref{eq:isoz0}) are not simply related. Indeed, the condition $\psi(0)=0$ leading to Eq. (\ref{eq:isoz0}) is not allowed for the minimal dispersion: it is not satisfied for any value of $\gamma$ in Eq. (\ref{eq:minBC}). 

At the level of boundary conditions, we may note that for large $\lambda$ the boundary conditions obtained from Eq. (\ref{eq:isonu}) reduce to Eq. (\ref{eq:minBC}) for $\delta=0, \alpha = \pm\pi/2$, and $m_0 = -\gamma -\alpha$. A straight Fermi arc is obtained for the minimal dispersion only when $\gamma = \pm\pi/2$, which corresponds to either $\delta=m_0=0, \alpha=-\gamma$ or $\delta=0, m_0=\mp\pi, \alpha=\gamma$. In any case, according to Eq. (\ref{eq:isolinFermiArc}), the isotropic dispersion for these boundary conditions also supports a straight Fermi arc.

\section{Conclusion}
We have studied the most general form of boundary conditions for a simple Hamiltonian of a Weyl semimetal with a single pair of Weyl nodes in the continuum limit. Our approach, based on self-adjoint extensions of the bulk Hamiltonian in bounded geometries, allows us to determine all physically possible boundary conditions and their respective surface spectra. These boundary conditions describe different physical interfaces at the boundary of the Weyl semimetal with other phases of matter. We also studied the role of bulk symmetries. In particular, we find that on the surface parallel to the direction of Weyl node separation, all physical boundary conditions break particle-hole symmetry and lead to chiral surface states. 

Using a minimal as well as an isotropic bulk dispersion, we determined physical surface spectra and found various types of spectra, including Fermi arcs, depending on the boundary conditions. On the surface normal to the direction of Weyl node separation, interesting gapped spectra arise for a wide family of boundary conditions. For example, a dispersion in the shape of a Mexican hat exists on this surface in a restricted region of the surface Brillouin zone. In the presence of interactions, or by engineering the local potential to induce coexisting particle and hole Fermi surfaces, such a spectrum could support novel correlated phases of matter at the surface. On the surface parallel to the Weyl node separation, we determined the general conditions giving the surface bound states and Fermi arcs. We found chiral surface bound states in restricted parts of the surface Brillouin zone; the chiral nature of the surface states is consistent with their topological nature.

Some interesting directions for future work are finding concrete models that realize given boundary conditions and studying the effects of interactions or potential profiles in cases of unusual surface state spectra. Another interesting direction is to study the general self-adjoint boundary conditions in other topological material, such as type-II Weyl semimetals.~\cite{SolGreWan15a,TchCivGoe17a}

\acknowledgements
This work was supported in part by the Indiana University REU program
through NSF grant PHY-1460882 (M.V.) as well as the NSF CAREER grant DMR-1350663 and the College of Arts and Sciences at Indiana University (B.S.). We also thank the hospitality of Aspen Center for Physics, supported by National Science Foundation grant PHY-1607611, where parts of this work were performed.

\appendix*
\section{Self-adjoint extensions}\label{app:SAE}
The distinction between self-adjoint and Hermitian operators is a subtlety that is usually ignored in introductions to quantum mechanics.~\cite{BonFarVal01a} One reason for this is that the two definitions coincide for non-singular Hamiltonians acting on systems that extend to infinity in all directions. When either of these conditions fail, the two are not the same, and a further analysis is needed. There is a strong physical basis for requiring that observables should correspond with self-adjoint operators. One key result comes from Stone's theorem,~\cite{ReeSim72a} which states that a transformation of the form $\exp(itA)$ is unitary for real $t$ if and only if $A$ is self-adjoint.

The difference between self-adjoint and Hermitian operators is a constraint on the functions that they can be defined on. All self-adjoint operators are Hermitian, but the converse is not true. Unbounded operators (derivatives in our case) have some properties with no analogue in finite-dimensional linear algebra. One such result is the Hellinger-Toeplitz theorem,~\cite{ReeSim72a} which states that unbounded, Hermitian operators cannot be defined on the entire Hilbert space, $\mathcal{H}$. 

With this context in mind, the definitions will be more transparent. Recall that an operator $A$ is Hermitian if
\begin{equation}
\langle \psi | A \phi \rangle = \langle A \psi | \phi \rangle
\end{equation}
for any elements $\psi$ and $\phi$ in the domain $\mathcal{D}(A)$ of $A$. The adjoint $A^{\dagger}$ of $A$ is defined through
\begin{equation}
\langle \psi | A \phi \rangle = \langle A^{\dagger} \psi | \phi \rangle
\end{equation}
for all $\phi\in\mathcal{D}(A)$ and $\psi$ in the adjoint domain $\mathcal{D}(A^{\dagger})$. A self-adjoint operator $A$ is a Hermitian operator with $\mathcal{D}(A) = \mathcal{D}(A^{\dagger})$. In the general case, the domains of $A$ and $A^\dagger$ are unrelated in the sense that one does not necessarily contain the other. However, when $A$ is Hermitian, the domain is always contained in the adjoint domain, that is $\mathcal{D}(A) \subseteq \mathcal{D}(A^{\dagger})$. We can interpret self-adjoint operators as maximally defined Hermitian operators. This suggests that there is a method of extending the domain of a Hermitian operator to promote it to a self-adjoint operator. von Neumann's theorem allows us to determine when this is true and classify all possible extensions.

\subsection{Deficiency indices and von Neumann's theorem}

The procedure outlined by von Neumann's theorem has the benefit that it may be applied to any Hermitian operator, $A$. In the first step, we compute its adjoint $A^{\dagger}$ and its domain $\mathcal{D}(A^{\dagger})$. The second step is to pick two values $+i\eta_+$ and $-i\eta_-$ laying in the upper and lower half of the complex plane respectively. With this choice, we solve the eigenvalue equations
\begin{equation}
A^{\dagger} \psi = \pm i\eta_{\pm} \psi
\end{equation}
and determine the eigenspaces $\mathcal{E}_{\pm} \subseteq \mathcal{D}(A^{\dagger})$ corresponding to $\eta_{\pm}$. An important result states that the deficiency indices $n_{\pm} = \dim \mathcal{E}_{\pm}$ do not depend on the choice of $\eta_{\pm}$. 

Now, von Neumann's deficiency index theorem states~\cite{ReeSim72a} that iff $n_+=n_-=n$, then $A$ has (infinitely many) self-adjoint extensions, $A_U$, parametrized by a $n\times n$ unitary map $U:\mathcal{E}_+\to \mathcal{E}_-$, and defined over the domain $\mathcal{D}(A_U) = \{ \phi_U=\phi + \psi_+ + U\psi_+ \}$ where $\phi\in\mathcal{D}(A)$ and $\psi_+\in\mathcal{E}_+ \subset \mathcal{D}(A^\dag)$, with the action $A_U \phi_U = A\phi + i\eta_+\psi_+ - i\eta_- U\psi_+$. In a bounded geometry, each self-adjoint extension corresponds to a boundary condition parameterized by $U$.

\subsection{Conserved-current formulation}
While von Neumann's deficiency index method is general, it is not physically transparent. A more intuitive picture of self-adjoint extensions is given by a conserved current defined, up to a constant, by~\cite{AhaOrtSer16a}
\begin{equation}
j_A = i \int^{x} \left[\psi^\dag(A \phi) - (A\psi)^\dag \phi\right]\hspace{-1.15mm}(\bar x)\: d\bar x,
\end{equation}
as in Eq.~(\ref{eq:SAEj}). In general, this current can be written as a sesquilinear form of $\psi$ and $\phi$ as $j=\nu[\psi]^\dagger J_A \nu[\phi]$ where $\nu[\psi]$ is a vector of the wavefunction and its derivatives and $J_A$ is a finite Hermitian matrix. In a semi-infinite geometry, the conservation condition $j_A(0)=0$ can then be written algebraically as
\begin{equation}
\nu_+[\psi(0)]^\dagger\nu_+[\phi(0)] = \nu_-[\psi(0)]^\dagger\nu_-[\phi(0)],
\end{equation}
where $\nu_+[\psi] \oplus \nu_-[\psi] := (d_+ \oplus d_-) T \nu[\psi]$ and $T$ is a unitary matrix that diagonalizes $J_A$ as $TJ_AT^\dagger = (d_+\oplus d_-)^2$, with $d_\pm = \pm d_\pm^\dagger$ diagonal. In this diagonal basis, $d_+$ ($d_-$) are the square roots of $J_A$ projected to the subspace of positive (negative) eigenvalues of $J_A$. This condition states that the inner product of $\nu_\pm$ must be preserved regardless of the states $\psi$ and $\phi$. Thus, the existence of self-adjoint extensions of $A$ is equivalent to the condition
\begin{equation}
\nu_+[\psi(0)] = U \nu_-[\psi(0)],
\end{equation}
for a unitary matrix $U$.~\cite{AhaOrtSer16a}

\bibliographystyle{physre}

\end{document}